\documentclass[showpacs,preprint,floatfix,amsmath,amssymb]{revtex4}

\usepackage{graphicx} 
\usepackage{dcolumn} 
\usepackage{bm} 
\begin{document} 
\title{Superconductivity in ropes of carbon nanotubes}
\author{M.~Ferrier$^1$, A.~De~Martino$^{2,}
\footnote{Corresponding author. Address: Institut f\"ur Theoretische Physik, 
Heinrich-Heine-Universit\"at, Universit\"atsstra{\ss}e 1, Geb\"aude 25.32,
D-40225 D\"usseldorf, Germany; E-mail: ademarti@thphy.uni-duesseldorf.de}
$, A.~Kasumov$^{1,3,} 
\footnote{Present address: RIKEN, Hirosawa 2-1, Wako, Saitama, 351-0198 Japan}$, 
S.~Gu\'eron$^1$, M. Kociak $^1$, R.~Egger$^2$ and H.~Bouchiat$^1$}
\affiliation{ 
$^1$Laboratoire de Physique des Solides, Associ\'e au CNRS, B\^atiment 510, 
Universit\'e Paris-Sud, F-91405 Orsay, France\\ 
$^2$Institut f\"ur Theoretische Physik, Heinrich-Heine-Universit\"at, 
D-40225 D\"usseldorf, Germany\\ 
$^3$Institute of Microelectronics Technology and High Purity Materials, 
Russian Academy of Sciences, Chernogolovka 142432 Moscow Region, Russia 
} 

\begin{abstract} 
Recent experimental and theoretical results on intrinsic superconductivity 
in ropes of single-wall carbon nanotubes are reviewed and compared. 
We find strong  experimental evidence for superconductivity when the distance 
between the normal electrodes is large enough. This indicates the presence of 
attractive phonon-mediated interactions in carbon nanotubes, which can 
even overcome the repulsive Coulomb interactions. 
The effective low-energy theory 
of rope superconductivity explains the experimental results on the 
temperature-dependent resistance below the transition temperature 
in terms of quantum phase slips. Quantitative agreement with only 
one fit parameter can be obtained. 
Nanotube ropes thus represent superconductors in an extreme 1D limit 
never explored before. 

\vspace*{.5cm}

\end{abstract}

\maketitle
\section{Introduction}
The hope to use molecules as the ultimate elementary 
building blocks for electronic 
circuits has motivated the quest to 
understand electronic transport in thinner 
and thinner wires, ideally with one or two conduction modes. 
However, a number of physical phenomena tend to drive one-dimensional (1D) 
metallic wires to an insulating state at low temperature. Carbon nanotubes, 
because of their special band structure, can escape such a fate and remain 
conducting over lengths greater than one micron down to very low 
temperature \cite{nts,dressel}. 
Moreover, transport through nanotubes has been shown 
to be quantum coherent \cite{Tans}. This 
is also demonstrated by the existence of strong supercurrents when 
individual nanotubes are connected 
to superconducting contacts \cite{Kasumov,morpurgo}. The observation 
of {\sl intrinsic superconductivity} 
in ropes of carbon nanotubes containing a few tens of tubes 
\cite{kociak,kasnew} 
is even more surprising and indicates 
the presence of attractive pairing interactions which overcome the strong 
repulsive interactions. This phenomenon is described in the present 
paper, both from the experimental and the theoretical point of view.
A single-wall nanotube (SWNT) is made of a single graphene plane 
wrapped into a cylinder. The Fermi surface of graphene reduces to two 
discrete points (usually denoted as $K$ and $K'$) at the corners of the first 
Brillouin zone \cite{wall}. 
As a result, depending on the diameter and helicity, which determine 
the boundary conditions of the electronic wave functions around the tube, 
a SWNT can be either semiconducting or metallic \cite{nts,dressel}. 
A metallic SWNT is characterized by just two conduction channels, 
low electronic density, Fermi velocity $v_F$ nearly 
as high as in copper, and long mean free path \cite{white}. 
These properties make them long sought-after realizations of 1D conductors. 
In one dimension, repulsive electron-electron interactions lead to an exotic 
correlated electronic state, the Luttinger liquid (LL) \cite{Luttinger,gogolin}. 
In a LL, collective plasmon-like 
excitations give rise to anomalies in the single-particle density of states, 
and long-range order cannot survive even at zero temperature. 
The low-energy theory of SWNTs \cite{Egger,Kane} predicts a metallic 
SWNT to constitute a realization of a four-channel LL, 
with channel index $a=c+,c-,s+,s-$ corresponding to 
total/relative charge/spin degrees of freedom. These arise due to the 
$K-K'$ degeneracy and the electronic spin. 
The interaction strength in a SWNT is then parameterized by a single 
dimensionless parameter $g$, where $g=1$ in the absence 
of interactions. For repulsive Coulomb interactions, 
this parameter is smaller than unity, with concrete estimates for 
SWNTs in the range $g\approx 0.2$ to 0.3 \cite{Egger,Kane}. For attractive 
interactions caused by phonon exchange, as long as 
retardation effects are negligible, one instead obtains a LL with 
$g>1$. In general, both effects have to be 
combined on equal footing. For superconductivity to occur, it 
appears to be necessary to have effectively $g>1$.

Experimental evidence for the validity of LL theory in a 
SWNT has been provided by measurements of the 
tunneling resistance diverging as a power 
law with temperature \cite{bockrath,yao}, from photoemission spectroscopy 
\cite{ishii}, and from transport properties of crossed SWNTs \cite{gao}. 
{}From these experimental results, values for the LL parameter in the range 
$g\approx 0.16$ to 0.3, consistent with theoretical expectations, 
were extracted. Such small values for $g$ correspond to pronounced 
repulsive interactions, and would imply that 
at very low temperature an insulating state is reached unless 
the material is extremely clean. 
The measurements in Refs.~\cite{bockrath,yao,gao} 
were done on individual nanotubes connected to the 
measuring leads through tunnel junctions. Because of the onset of Coulomb 
blockade \cite{Grabert}, the low-temperature small-voltage 
regime has not been explored in depth. 
We have developed a technique in which measuring pads are connected 
through low-resistance contacts to {\sl suspended}\ nanotubes \cite{Kasumov2}. 
Ropes and individual SWNTs connected to normal contacts using this 
technique exhibit only very weak temperature and bias dependence of the 
resistance down to 1K. More surprisingly, we have reported 
experimental evidence of intrinsic 
superconductivity below $0.5$~K in ropes, provided that the distance 
between the normal electrodes is large enough \cite{kociak,kasnew}. 
In this paper, we discuss the 1D character of the transition and 
the physical parameters that govern this transition, such as the 
length of the rope, the number of metallic SWNTs in the rope, 
the intertube couplings, disorder, and so on. 
Below we summarize both our experimental results \cite{kociak,kasnew} 
and the low-energy theory describing the superconducting 
state in ropes proposed by two of us \cite{ademarti,ademarti2}. 
Fortunately, the measured low-temperature data for the resistance allow 
to perform detailed tests of the theory.
In this theory, the rope is modelled as an array of $N$ metallic 
SWNTs with effectively attractive 
intratube interactions, coupled together by Cooper-pair hopping. 
Attractive phonon-mediated interactions may overcome the 
Coulomb repulsion in a sufficiently thick rope, 
leading to a LL parameter $g>1$. 
The dominant 1D fluctuations on individual SWNTs then cause the incipient 
formation of singlet Cooper pairs. 
Superconductivity of the rope is finally stabilized by Cooper 
pair hopping between the tubes (Josephson coupling), see also 
Refs.~\cite{gonzalez1,gonzalez2,gonzalez3}. 
Since typical elastic mean free paths 
in metallic SWNTs may exceed 1~$\mu$m \cite{nts,white}, intratube disorder is 
completely neglected. However, disorder due to the random distribution 
of tube chiralities in the rope, where only $1/3$ of the SWNTs is expected to 
be metallic \cite{nts,dressel}, is taken into account in the following way. 
First, due to momentum-conservation arguments, 
it strongly suppresses single-particle hopping between adjacent 
SWNTs \cite{kane,gonzalez1}, which is thus neglected henceforth. 
Second, we introduce a matrix $\Lambda$, whose elements 
$\Lambda_{ij}$ represent the 
Josephson couplings between the $i$th and the $j$th tube, where 
$i,j=1,\dots,N$. 
In order to simulate the random distribution of tube chiralities, 
$\Lambda$ should be drawn from an appropriate random distribution. 
Typically, $\Lambda_{ij} \approx a_0 
t_{\perp}^2/ \Delta E$ when metallic tubes are nearest neighbors,
and zero otherwise. Here, $a_0=0.24$~nm is the lattice
spacing, $t_{\perp}$ is the transverse intertube hopping energy,
 and $\Delta E$ is the typical energy band
 spacing within one tube \cite{gonzalez2}.
However, detailed information about $\Lambda$ is not needed 
in the low-energy regime, and general results can be derived 
for a fixed but unspecified $\Lambda$. 
For this model, the effective action for the proper 
order parameter allows to identify 
a mean-field transition temperature $T_c^0$. 
For $T<T_c^0$, the amplitude of the order parameter is finite, 
but due to the reduced dimensionality, phase fluctuations 
may still destroy superconductivity \cite{Tinkham}. 
Such fluctuations are shown to indeed cause a 
depression of the true transition temperature $T_c$ 
below the mean-field value, which can be linked to the 
proliferation of quantum phase slips (QPSs). 
A QPS is a topological vortex-like excitation of 
the superconducting phase field, which only exists in 
1D superconductors \cite{Tinkham}. 
In addition to the $T_c$ depression, QPSs 
produce a finite sub-$T_c$ linear resistance apart from 
the usual temperature-independent contact resistance. 
This effect is 
indeed observed experimentally, and can be compared in a 
quantitative way to theory. Our theory makes detailed predictions 
about the temperature dependence of this resistance, where almost all 
free parameters can be determined independently. There is essentially 
only one free (dimensionless) fit parameter, 
which should have a value close to unity. This turns indeed out to be 
the case.

This paper is organized as follows. 
In Sec.~\ref{sec2}, we summarize and discuss 
experimental results on intrinsic superconductivity in 
ropes of carbon nanotubes. 
In Sec.~\ref{sec3}, the effective low-energy theory 
of rope superconductivity is reviewed. 
In Sec.~\ref{sec4}, the theoretical predictions are compared 
with the experimental results. The rather good agreement found there 
supports the notion that ropes represent 
1D superconductors in the few-channel limit, where 
QPSs can be experimentally observed in a clear manner from the 
temperature-dependent resistance below $T_c$.

\section{Experimental evidence for rope superconductivity} \label{sec2}
\begin{table}[bp] 
\begin{tabular}{|c|c|c|c|c|c|c|c|} 
\hline 
&L & N & $R_{290K}$ & $R_{4.2K}$&T*&$I_c$&$I_{c}$$^*$ \\ \hline 
$R1_{PtAu}$ & $2~\mu m$ & 350 & 10.5 $k\Omega$ & 1.2 $k\Omega$ &140 mK & 
0.1 $\mu A$ &0.36 $\mu A$ \\ \hline 
$R2_{PtAu}$ &$1~\mu m$& 350 & 4.2$k\Omega$ & 9.2$k\Omega$&550 mK&0.075 $\mu 
A$&3 $\mu A$\\ \hline 
$R3_{PtAu}$&$0.3 \mu m$& 350 &400 $\Omega$&450 $\Omega$&*&*&*\\ \hline 
$R4_{PtAu}$&$1 \mu m$&45&620 $\Omega$&620 $\Omega$&120 mK&*&$0.1~\mu A$ \\ \hline 
$R5_{PtAu}$&2 $\mu m$&300&16 $k\Omega$&21 $k\Omega$&130 mK&20 $n A$&0.12 $\mu A$\\ \hline 
$R6_{PtAu}$&0.3 $\mu m$&200&240 $\Omega$&240 $\Omega$&*&*&*\\ \hline 
\end{tabular} 
\caption{ \label{table1} Summary of the characteristics of 
six ropes mounted on $Pt/Au$ contacts. 
$T^*$ is the transition temperature 
below which the resistance starts to drop, $I_c$ is the 
current at which the first resistance increase occurs, and $I_{c}^*$ is the 
current at which the last resistance jump occurs.} 
\end{table}
\begin{figure}[h] 
\begin{center} 
\includegraphics[clip=true,width=9.5cm]{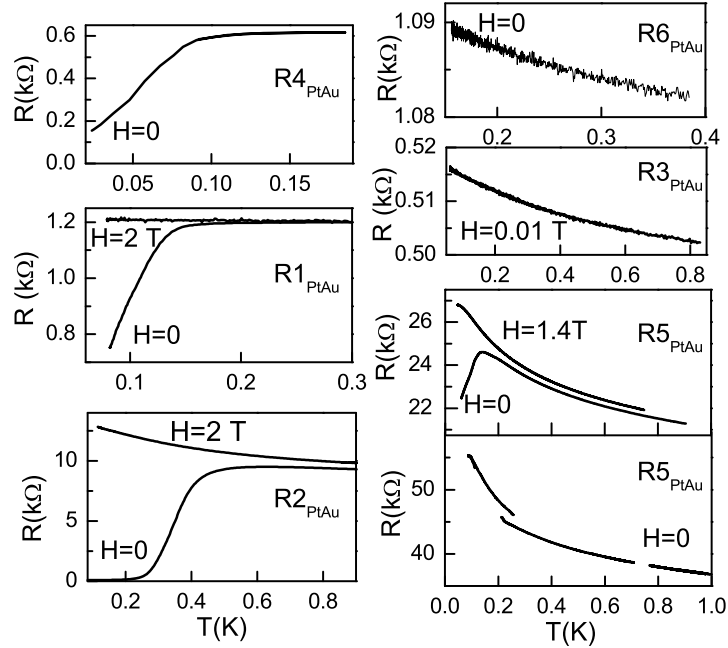} 
\end{center} 
\caption{ \label{figurert} 
Resistance as a function of temperature for the six samples described 
in Table \ref{table1}, both for zero and large magnetic fields.} 
\end{figure}
\begin{figure}[h] 
\begin{center} 
\includegraphics[clip=true,width=8cm]{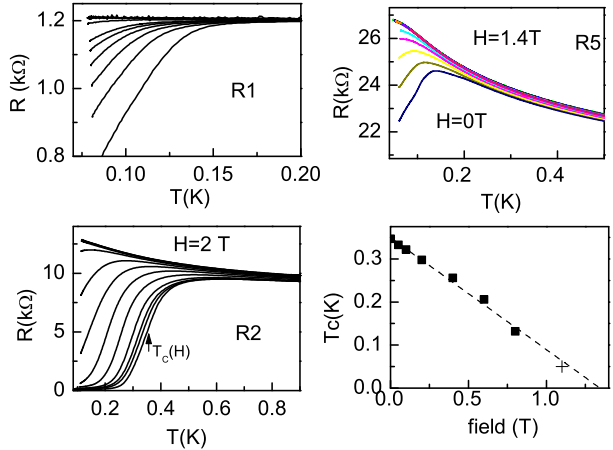} 
\end{center} 
\caption{ \label{figurerth} 
Resistance as a function of temperature for samples $R1,2,5_{PtAu}$ showing 
a transition. The resistance of $R1$ is measured in magnetic fields of 
$\mu_{0}H$= 0, 0.02, 0.04, 0.06, 0.08, 0.1, 0.2, 0.4, 0.6, 0.8 and 1 T 
from bottom to top. The resistance of $R2$ is taken at 
$\mu_{0}H$=0, 0.05, 0.1, 0.2, 0.4, 0.6, 0.8, 1, 1.25, 1.5, 1.75, 2, 2.5 T 
(from bottom to top), that of $R5$ at $\mu_{0}H$=0, 0.1, 0.2, 0.3, 0.5, 1.4 T 
(from bottom to top). Bottom right: $T_c(H)$ for $R2$.} 
\end{figure}
In this section, we review experimental results from Ref.~\cite{kasnew} 
reporting evidence for intrinsic superconductivity in ropes of carbon nanotubes. 
We start with a discussion of the 
low-temperature (below 1~K) transport regime of suspended 
ropes of SWNTs connected to normal electrodes. The electrodes are 
trilayers of sputtered $Al_{2}O_{3}/Pt/Au $ of respective thickness 5, 3 
and 200~nm. They do not show any sign of superconductivity down to 50 mK. 
As is shown in Fig.~\ref{figurert}, different behaviors are observed 
for the temperature dependence of the linear resistance. The 
resistance of short samples whose
length is of the order of $0.3$~$\mu$m ($R3_{PtAu}$ and $R6_{PtAu}$) increases 
weakly and monotonously as temperature is reduced, whereas the resistance of 
samples longer than $1$~$\mu$m  ($R1,2,4,5_{PtAu}$) drops over a relatively broad temperature 
range, starting below a transition temperature $T^*$ between 0.4 and 0.1 K, 
see Table \ref{table1}. The resistance 
of $R1_{PtAu}$ is reduced by 30\% at 70~mK, and that of $R4_{PtAu}$ by 75\% 
at 20~mK. In both cases, no inflection point in the temperature dependence 
is observed. On the other hand, the resistance of $R2_{PtAu}$ decreases by 
more than two orders of magnitude, and reaches a constant value below 100~mK, 
namely $R_{r}=74$~$\Omega$. This residual resistance $R_r$ 
is interpreted as a contact resistance which must be present 
even in the superconducting phase. The contact resistance arises because 
only a finite number $N$ of metallic SWNTs is coupled to the normal-conducting 
pads. Since each metallic SWNT has two spin-degenerate conduction 
channels, 
\begin{equation} \label{res} 
R_r=R_Q/2N , \quad 
R_{Q}=h/2e^2=12.9~\rm{k}\Omega. 
\end{equation} 
Since the residual resistance can be experimentally determined 
quite accurately, at least for the two samples $R2$ and $R4$, 
the number $N$ can be obtained directly using Eq.~(\ref{res}). 
This number is an important parameter for the theory described in 
Sec.~\ref{sec3}.

\begin{figure}[hbt] 
\begin{center} 
\includegraphics[clip=true,width=9cm]{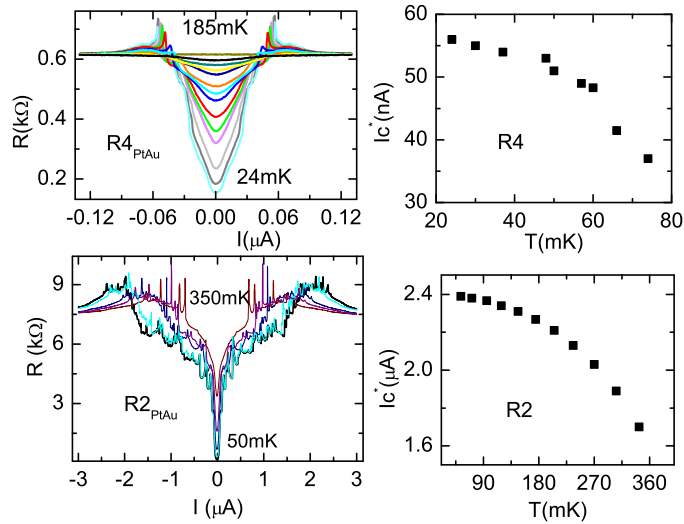} 
\end{center} 
\caption{ 
\label{figuredivt} 
Differential resistance of $R2 _{PtAu}$ and $R4 _{PtAu}$ at different 
temperatures. Right panel: Temperature dependence of $I_c^*$, the 
current at which the last resistance jumps occur in the $dV/dI$ curves.} 
\end{figure}

\begin{figure}[hbt] 
\begin{center} 
\includegraphics[clip=true,width=8cm,height=7cm]{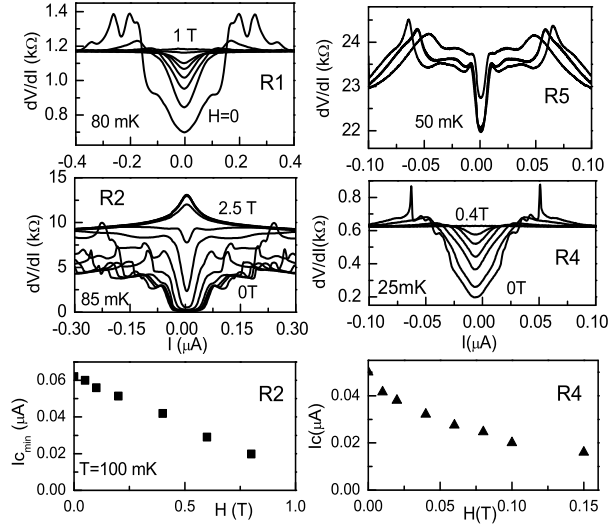} 
\end{center} 
\caption{ 
Differential resistance as a function of current for samples $R1,2,4,5_{PtAu}$ 
in different applied fields. Sample $R1$: Fields are 0, 0.02, 0.04, 0.06, 0.08, 
0.1, 0.2 and 1 T. Sample $R2$: Fields are 0, 0.2, 0.4, 0.6, 0.8, 1, 1.25, 1.5, 
1.75, 2, and 2.5 T. Sample $R5$: Fields are 0.02, 0.04, 0.06, 0.08 T. 
Sample $R4$: Fields are 0, 0.02, 0.06, 0.1, 0.15, 0.2 and 0.4 T. Bottom: 
Field dependence of $I_{c}$ for samples $R2_{PtAu}$ and $R4_{PtAu}$. 
Note the linear behavior.\label{figuredivh}} 
\end{figure}

The low-temperature drop of the resistance 
below $T^*$ disappears when increasing the 
magnetic field. For all samples, a critical field can be defined, above 
which the normal-state resistance is recovered. 
As shown in Fig.~\ref{figurerth}, 
this critical field decreases linearly with temperature, very similar to what 
is seen in SWNTs and ropes connected to superconducting contacts 
\cite{Kasumov,morpurgo}. We define a 
critical field $H_c$ as the extrapolation of $H_c(T)$ to 
zero temperature, see Fig.~\ref{figurerth}. 
Above the critical field, the resistance increases with 
decreasing temperature, 
similar to ropes $R3$ and $R6$, and becomes independent of magnetic field. 
Figures \ref{figuredivt} and \ref{figuredivh} show that in the temperature 
and field range where the 
linear resistance drops, the differential resistance 
is strongly current-dependent, with lower resistance at low current. These 
data suggest that the ropes $R1, R2,$ and $R4$ are 
superconducting. Although the 
experimental curves for $R2_{PtAu}$ look similar to those of SWNTs connected 
to superconducting contacts \cite{Kasumov}, there are major differences. 
In particular the $V(I)$ and $dV/dI(I)$ curves found here do not show any 
supercurrent because the contacts are normal metals, implying 
a finite residual resistance.

\begin{figure} 
\begin{center} 
\vspace{.8 cm} 
\includegraphics[width=9cm]{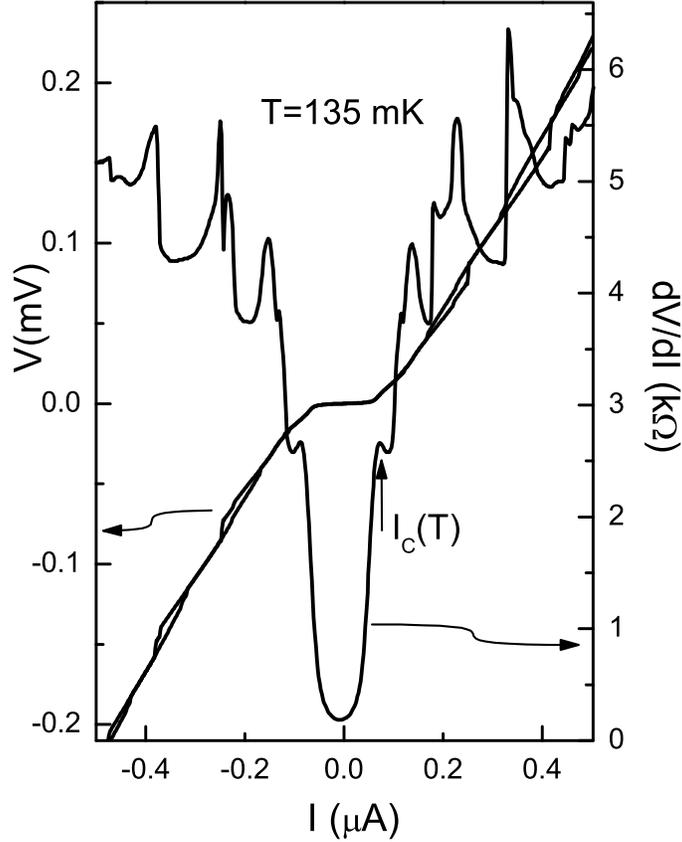} 
\end{center} 
\caption{ 
\label{pslips} 
$V(I)$ and $dV/dI(I)$ curves for sample $R2_{PtAu}$. 
Note the hysteretic behavior in $V(I)$ at each peak 
in the $\frac{dV}{dI}(I)$ curve. }
\end{figure}
\begin{figure}[h] 
\begin{center} 
\includegraphics[width=5cm]{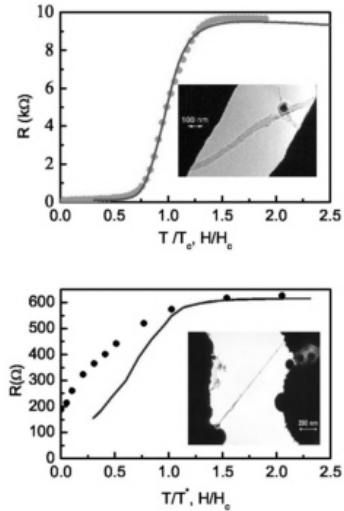} 
\end{center} 
\caption{ 
\label{Pt24} 
Resistance as a function of temperature (continuous line) and magnetic 
field (scatter points) for samples $R2_{PtAu}$ and $R4_{PtAu}$. 
Insets: TEM micrographs of the samples.} 
\end{figure}

The observed jumps in the differential resistance as the current is 
increased, see Fig.~\ref{pslips}, are similar to the behavior observed  in long narrow 
superconducting metal wires in the very 
vicinity of the transition temperature. However,
in the present case these jumps are observed
 down to very low temperature \cite{Giordano}. 
For sample $R2$, the differential resistance at low current remains equal 
to $R_{r}$ up to 50 nA, where it strongly rises but does not recover 
its normal-state value until 2.5~$\mu$A. The jump in resistance at the first 
step corresponds approximately to the normal-state resistance of a length 
$\xi$ of sample $R2$ ($R_\xi$), where $\xi$ is the superconducting 
coherence length estimated from $\xi=\sqrt{\hbar D/ \Delta}$ 
where $\Delta $ is the BCS gap related to the transition
 temperature $T^*$ and $D$ the diffusion constant
describing transport in the rope in the normal state.
 Each peak corresponds to a hysteretic feature in 
the $V-I$ curve, see Fig.~\ref{pslips}. These jumps are identified as 
phase slips \cite{Tinkham,Giordano,meyer}, which reflect the occurrence of 
normal regions located around defects in the sample. Such phase slips 
could be thermally activated phase slips (TAPSs), leading to a 
roughly exponential decrease of the resistance instead of 
a sharp transition. At sufficiently 
low temperature and voltage, instead of the 
TAPSs, quantum phase slips (QPSs) of typical size $\xi$ are expected to dominate. 
The competition between TAPS and QPS processes is addressed briefly in 
Sec.~\ref{sec3} below. 
In sample $R2$, the current at which the first resistance jump occurs 
(60 nA, see Fig. \ref{figuredivh}) is close to the critical 
current expected theoretically for 
a diffusive superconducting wire \cite{klap}, 
\[ 
I_{c}=\Delta_{2} /eR_{\xi} \approx 20~{\rm nA}, 
\] 
with gap $\Delta_{2}=85$~$\mu$eV.
On the other hand, 
the current at which the last resistance jump occurs (2.4 $\mu$A, 
see Fig.~\ref{figuredivt}) is close to the critical current of 
a ballistic superconducting wire with the same number of conducting channels 
\cite{Tinkham}, 
\[ 
I_{c}^{*}=\frac{\Delta_{2}}{eR_{r}} \approx 1~\mu{\rm A}. 
\] 
Before analyzing the data further, we wish to emphasize that 
this is the first observation of superconductivity in wires with 
$N<100$ conduction channels. 
Earlier experiments in nanowires \cite{Giordano,dynes,tinkham2} dealt 
with at least a few thousand channels. We therefore expect a strong 1D 
behavior for the transition. In particular, the broadness of the 
resistance drop with temperature is linked to large fluctuations of the 
superconducting order parameter as expected in one dimension. 
An important parameter is the number of tubes in the rope. If there 
are only a few tubes in the rope, the system is very 
close to the strict 1D limit, and the 
transition is very broad. Comparing the two ropes in Fig.~\ref{Pt24}, 
it is clear that the transition, both in temperature and magnetic field, 
is much broader in sample $R4_{PtAu}$ with only $\approx 45$ tubes 
than in $R2_{PtAu}$ with $\approx 350$ tubes. 
Moreover, there is no inflection 
point in the temperature dependence of the resistance in the thinner 
rope, typical of a strictly 1D behavior. We also expect a stronger 
screening of the repulsive Coulomb 
interactions in the thick rope, 
which is also in favor of superconductivity. 
In the following, when comparing to theoretical predictions, we will 
have to take into account several essential features, e.g., 
the influence of the normal contacts, 
the finite length of the samples compared to 
relevant mesoscopic and 
superconducting scales, the effects of disorder, 
and the role of intertube couplings. 

\section{Effective low-energy theory of rope superconductivity} 
\label{sec3}
\begin{figure} 
\begin{center} 
\includegraphics[clip=true,width=7cm,height=6cm]{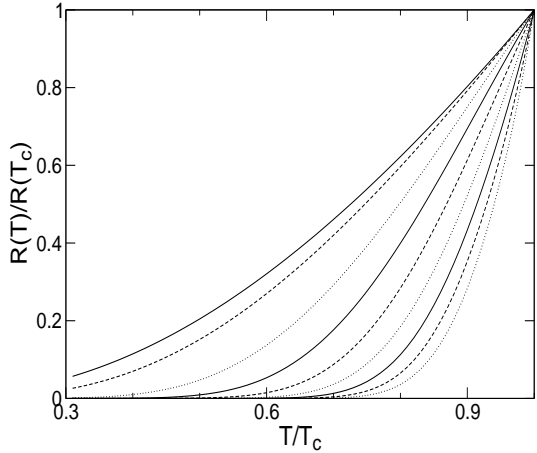} 
\end{center} 
\caption{\label{figevo} 
Temperature-dependent resistance $R(T<T_c)$ 
predicted by Eq.~(\ref{resis}) 
for $\nu=1$ and different $N$. The smaller is $N$, the broader is 
the transition. {}From the leftmost to the rightmost curve, 
$N=4,7,19,37,61,91,127,169,217$. 
} 
\end{figure}

In this section we summarize the main features of the recently proposed 
effective low-energy theory of intrinsic superconductivity 
in carbon nanotube ropes. For technical details, we refer the 
reader to the original publication \cite{ademarti2}. 
The basic ingredients of the model have been discussed in the 
Introduction, see also Refs.~\cite{gonzalez1,gonzalez2,gonzalez3,kane}. 
Within the standard bosonization approach \cite{gogolin}, 
the model is described by the Euclidean action 
\begin{equation}\label{ea} 
S = \sum_{j=1}^N S_{\rm LL}^{(j)} - \sum_{jk} 
\Lambda_{jk}\int d x d\tau \, {\cal O}^\ast_j {\cal O}_k^{} , 
\end{equation} 
where $-L/2<x<L/2$ is the spatial 1D coordinate along the individual 
SWNTs for rope length $L$, and 
$0\leq \tau < 1/T$ is imaginary time (we put $\hbar=k_B=1$ in intermediate 
steps). 
The first term describes the metallic 
tubes in the rope as $N$ uncoupled identical four-channel Luttinger liquids 
\cite{Egger,Kane,ademarti}, 
\begin{equation}\label{bosac} 
S_{\rm LL} = \int dx d\tau \sum_{a=c\pm,s\pm} 
\frac{v_a}{2 g_a} \left[ (\partial_\tau \varphi_{a}/v_a)^2 + 
(\partial_x \varphi_{a})^2 \right]. 
\end{equation} 
Here the boson fields $\varphi_a(x,\tau)$ 
(and associated dual fields $\theta_a$)
describe the collective total/relative charge/spin plasmon-like 
excitations. 
The interaction parameter for the total charge mode $g_{c+}\equiv g$ 
is determined by the combined effect of Coulomb repulsion 
and phonon-mediated attractive interactions. If Coulomb interactions 
are screened off, e.g., by the other SWNTs in a thick rope \cite{gonzalez3}, 
effectively attractive interactions are possible. 
In what follows, we assume $g>1$, where $g\approx 1.3$ 
has been estimated for (10,10) SWNTs with good screening \cite{ademarti}. 
The interaction parameters in the neutral channels are 
practically not affected by interactions, 
$g_{c-,s+,s-}=1$. The velocities $v_a$ in Eq.~(\ref{bosac}) 
are given by $v_a=v_F/g_a$, where $v_F=8\times 10^5$~m$/$sec 
is the Fermi velocity. 
The second term in Eq.~(\ref{ea}) describes the intertube couplings 
in the form of Cooper-pair hopping. 
The combined effects of random tube chiralities and 
attractive electron-electron interactions drive the incipient 
formation of singlet Cooper pairs on individual SWNTs. 
The Cooper pair operator ${\cal O}$ in Eq.~(\ref{ea}) is in 
bosonized language expressed as \cite{egger98} 
\begin{equation}\label{orderpar} 
{\cal O}= \frac{1}{\pi a_0} 
\cos[\sqrt{\pi} \theta_{c+}] \cos[\sqrt{\pi} \varphi_{c-}] 
\cos[\sqrt{\pi} \varphi_{s+}] \cos[\sqrt{\pi} \theta_{s-}] - 
(\cos\leftrightarrow \sin), 
\end{equation} 
where we identify the UV cutoff necessary in the bosonization 
scheme with the lattice constant $a_0$.
In order to investigate the physical properties of the system 
described by the action (\ref{ea}), 
approximations are necessary. Since in 1D systems fluctuations 
are strong, we can expect that the order parameter amplitude remains small 
over a wide temperature range, and a Ginzburg-Landau-type expansion 
should be accurate. To that end, 
we first decouple the Josephson terms in 
Eq.~(\ref{ea}) by performing a 
Hubbard-Stratonovich transformation \cite{nagaosa1}. 
This introduces a complex field $\Delta_i(x,\tau)$, 
which acts as the superconducting 
order parameter. The partition function for the original system can then be 
expressed as a functional integral over the fields $\Delta_i$, with 
an effective action formally defined as a functional integral 
over the LL boson fields. The latter functional integral 
cannot be performed analytically, and one has to resort to approximate methods. 
A systematic approach proceeds via cumulant expansion, where 
the small expansion parameter is 
$|\Delta|/2{\pi}T$, and one has to keep terms (at least) up to quartic order 
\cite{nagaosa1}. After perfoming 
a gradient expansion, justified for slow temporal and spatial variations 
of the order parameter, i.e., in the low-energy long-wavelength regime 
of primary interest here, 
one finally obtains a quantum Ginzburg-Landau (GL) action, 
\begin{equation} 
\label{hs2} 
S = \int dx d\tau\left \{\sum_{j=1}^N \left[ 
\left( \Lambda_1^{-1} - A \right) |\Delta^{}_j|^2 
+ B |\Delta^{}_j|^4 \right] + 
C |\partial_x \Delta_j^{}|^2 + D |\partial_\tau \Delta_j^{}|^2 
+ \sum_{jk}\Delta^\ast_j V_{jk} \Delta^{}_k \right\} , 
\end{equation} 
with positive temperature-dependent coefficients $A,B,C,D$ 
and a real symmetric positive-definite matrix $V_{ij}$ defined in terms 
of the Josephson matrix $\Lambda$, see Ref.~\cite{ademarti2} for details. 
Furthermore, $\Lambda_1$ is the largest 
eigenvalue of $\Lambda$. The GL 
coefficients can be computed analytically from this expansion, 
and are expressed in terms of the microscopic model parameters \cite{ademarti2}. 
Note that we keep the imaginary-time dependence of $\Delta_i(x,\tau)$, 
which is essential for what follows. Thereby quantum fluctuations 
are fully accounted for, in contrast to 
standard static GL theory \cite{Tinkham}.
The coefficient $A(T)$ is found to grow as $T$ decreases, 
and hence a mean-field critical temperature follows from the condition 
$A(T^0_c)=\Lambda^{-1}_1$. The result is 
\begin{equation}\label{tc} 
T_c^0 = c_0 \frac{\hbar v_F}{ k_B a_0} ( \Lambda_1 /\hbar v_F )^{2g/(g-1)} , 
\end{equation} 
with $c_0$ a dimensionless prefactor of order unity. 
$T_c^0$ exhibits a dependence on $N$ and the connectivity of the 
Josephson matrix through the eigenvalue $\Lambda_1$. 
For large $N$, $\Lambda_1(N)$ saturates, and Eq.~(\ref{tc}) 
approaches the bulk transition temperature. 
Using $\Lambda_1$ estimates from Ref.~\cite{gonzalez2} 
and typical $N$ from Table \ref{table1}, we find 
$T_c^0\approx 0.1$ to 1~K. A precise estimate is 
difficult to give because the Josephson matrix is in 
general unknown, and due to the typically large exponent in 
Eq.~(\ref{tc}), $T_c^0$ 
depends very sensitively on $\Lambda_1$.
For $T<T_c^0$, it is convenient to 
adopt an amplitude-phase representation of the order 
parameter, $\Delta_j=|\Delta_j|\exp[i\phi_j]$. 
The amplitude of the order parameter field is then 
finite, with a gap for fluctuations around the mean-field value. 
This mean-field value can be directly calculated from the 
saddle-point equation for the action (\ref{hs2}). 
The numerical solution to this equation shows 
that the GL expansion parameter $|\Delta|/2\pi T$ indeed remains small down 
to very low temperatures, and the use of GL theory is self-consistently 
justified \cite{ademarti2}. Due to the mass gap for amplitude fluctuations, 
the amplitudes can then be fixed to their mean field value. 
Moreover, for $N<100$, transverse fluctuations are 
negligible both regarding the amplitude, as follows from 
the numerical solution of the saddle-point equations, as well as concerning 
the phase, which follows from scaling dimension arguments. 
Therefore, one finally arrives at a standard 
Gaussian action governing the dynamics of 
the superconducting phase \cite{Tinkham}, 
\begin{equation}\label{finala} 
S=\frac{\mu}{2\pi}\int dx d\tau \left[ 
c_s^{-1}(\partial_\tau\phi)^2 + c_s (\partial_x \phi)^2 \right], 
\end{equation} 
where $\phi_j=\phi(x,\tau)$ is 
equal on all tubes. The superconductor's phase is the relevant 
fluctuation mode in a 1D system. Furthermore, 
$c_s$ is the Mooij-Sch\"on mode velocity \cite{mooij}, which here 
is of order $v_F$, and the 
dimensionless rigidity follows in the form 
\begin{equation} \label{mu1} 
\mu(T) = N \nu \left[ 1 - (T/T_c^0)^{(g-1)/2g} \right] , 
\end{equation} 
where $\nu \approx 1$. 
Equation (\ref{mu1}) for the temperature-dependent phase stiffness 
is one of the central results in Ref.~\cite{ademarti2}. 
The value of $\nu$ is difficult to compute in a very precise way. 
In particular, disorder and dissipation effects neglected in our model 
tend to decrease it \cite{zaikin1,zaikin2}. 
Therefore, $\nu$ is considered below as 
a fit parameter when comparing with experimental data. 
In fact, $\nu$ turns out to be essentially the only free fit parameter, 
where internal consistency of the theory constrains $\nu$ to 
be of order unity.
In the 1D system described by the action (\ref{finala}), 
vortex-like topological excitations (quantum phase slips of size $\xi$) can destroy 
superconductivity and give rise to a broad resistive transition. 
This occurs via a Kosterlitz-Thouless transition 
\cite{chaikin,zaikin1,zaikin2}. 
For $\mu(T)> 2$, phase slips are confined into 
neutral pairs, and superconductivity (in the 1D sense of quasi-long-range 
order) survives the phase fluctuations governed by Eq.~(\ref{finala}). 
However, for $\mu(T)<2$, phase slips proliferate, 
the phase stiffness is renormalized to zero, and the system is driven 
to a normal phase. The true transition temperature then follows 
from the condition $\mu(T_c) =2$, which gives 
\begin{equation}\label{tc1} 
\frac{T_c}{T_c^0} = \left[1-\frac{2}{\nu N}\right]^{2g/(g-1)}. 
\end{equation} 
Equation~(\ref{tc1}) implies that QPSs can lead to 
a sizeable depression of the mean-field critical temperature for 
$N<100$.
In Ref.~\cite{ademarti2} the 
relative contribution of thermally activated and quantum 
phase slips has been estimated. Using results from 
Ref.~\cite{zaikin2} it has been shown that the crossover 
temperature between the two regimes is of the order of $T_c^0$, 
which implies that for $T<T_c$, with $T_c<T_c^0$, only QPSs 
significantly affect the resistance below $T_c$. 
Indeed, even in the superconducting phase with $\mu>2$, 
QPS fluctuations produce a sizeable 
resistance. Under a small constant current bias, this resistance 
can be computed from the voltage drop $V$ associated to the 
occurrence of phase slips via the Josephson relation, see 
also Ref.~\cite{zaikin1}. 
The final result for the linear resistance $R(T<T_c)$ 
for arbitrary (but larger than $\xi$) rope length $L$ and 
thermal length $L_T=\hbar c_s/\pi k_B T$ is \cite{ademarti2} 
\begin{equation}\label{resis} 
\frac{R}{R_Q}= \left( \frac{\pi y \Gamma(\mu/2)}{ \Gamma(\mu/2+1/2)} 
\right)^2 \frac{\pi L}{2\kappa} \left(\frac{L_T}{\kappa}\right)^{3-2\mu} 
\int_{0}^\infty du \frac{2/\pi}{1+u^2} 
\left|\frac{\Gamma(\mu/2+iu L_T/2L)} {\Gamma(\mu/2)} \right|^4 , 
\end{equation} 
where $R_Q$ is the resistance quantum, see Eq.~(\ref{res}), $\Gamma(x)$ denotes 
the Gamma function, and 
$y$ and $\kappa$ are the QPS fugacity 
and core size, respectively. 
Equation (\ref{resis}) has been obtained from 
perturbation theory in the fugacity $y$, and thus only 
holds for a dilute QPS gas. This assumption breaks 
down in the vicinity of $T_c$, where QPSs proliferate 
and cause the Kosterlitz-Thouless transition. Equation (\ref{resis}) 
is therefore only reliable well below $T_c$, and cannot be used to describe 
the resistance saturation around $T=T_c$. 
For $L/L_T\gg 1$, the $u$-integral approaches unity, 
and hence $R\propto T^{2\mu-3}$, 
while for $L/L_T\ll 1$, dimensional scaling arguments 
give $R\propto T^{2\mu-2}$ \cite{zaikin1}. In Refs.~\cite{kociak,kasnew} 
typical lengths were $L\approx 1~\mu$m, which puts one into the 
crossover regime $L_T\approx L$. 
Inspection of Eq.~(\ref{resis}) also shows that, as expected, 
the transition becomes broader and broader when the number 
of tubes decreases. This is illustrated in Fig.~\ref{figevo}, 
where the theoretical resistance curves are depicted for various 
values of $N$, taking $\nu=1$ and $g=1.3$.

\section{Comparison with experiments} 
\label{sec4}
\begin{figure} 
\begin{center} 
\includegraphics[clip=true,width=7cm,height=6cm]{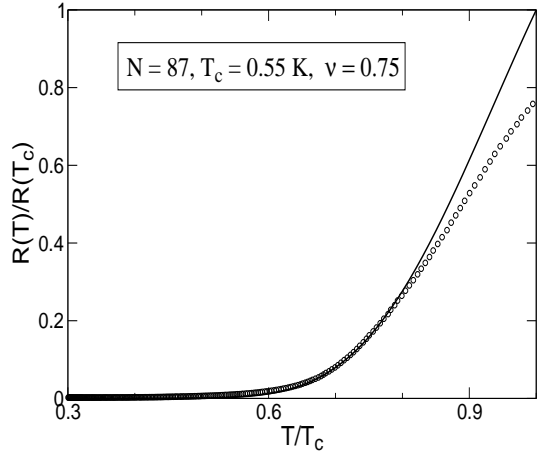} 
\end{center} 
\caption{\label{figR2} 
Temperature dependence of the linear resistance below $T_c$ for 
sample $R2$. Open circles denote experimental data 
(with subtracted residual resistance corresponding to $N=87$), 
the curve is the theoretical result for $\nu=0.75$. 
} 
\end{figure}

\begin{figure}[t] 
\begin{center} 
\includegraphics[clip=true,width=7cm,height=6cm]{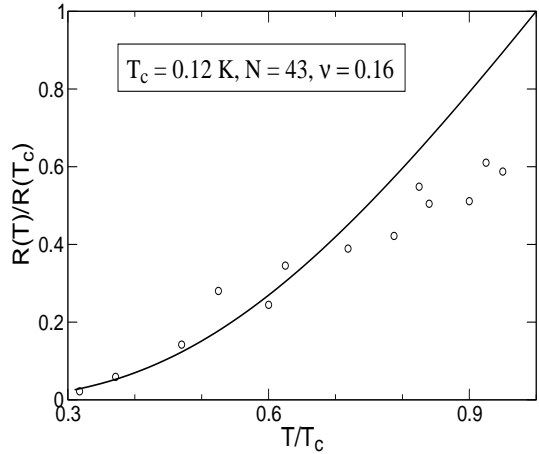} 
\end{center} 
\caption{\label{figR4} 
Same as Fig.~\ref{figR2}, but for sample $R4$ with $N=43$ and 
$\nu=0.16$. 
} 
\end{figure}

In this section we compare the theoretical result (\ref{resis}) for the 
temperature-dependent resistance below $T_c$ with the experimental data. 
We focus on samples $R2$ and $R4$, where the resistance has been 
measured down to quite low temperatures, and a meaningful comparison 
is possible, see Figs.~\ref{figR2} and \ref{figR4}. 
In this comparison, it has to be borne in mind that Eq.~(\ref{resis}) 
does not take into account the normal contacts. These cause a 
contact resistance (\ref{res}) which we subtract from the experimental 
data when comparing to Eq.~(\ref{resis}). In addition, the experimentally 
measured residual resistance fixes the value of $N$ taken in the 
respective comparison. 
Moreover, Eq.~(\ref{resis}) does not take into account the 
possible destruction of superconductivity by the normal contacts. 
Indeed, investigation of the proximity effect at high-transparency NS 
interfaces has shown that superconductivity resists the presence of 
normal contacts only if the length of the superconductor is much greater 
than its coherence length $\xi$ \cite{Belzig}. This is probably the reason 
why superconductivity is only observed in the longest ropes. 
Finally, Eq.~(\ref{resis}) only applies to temperatures well 
below $T_c$, and in particular does not capture quasiparticle 
effects or phonon backscattering. The transition to the normal-state 
resistance is not described at this level of theory.
Besides $N$ and the LL parameter, which is taken to be $g=1.3$, 
another important parameter appearing in the expression 
for the resistance is the critical temperature $T_c$. In principle 
this could be computed once the eigenvalue $\Lambda_1$ 
of the coupling matrix $\Lambda$ is known. However, as discussed above, 
to obtain a reliable estimate for $\Lambda_1$ is very difficult. In order 
to circumvent this problem, we remark that it is natural to identify $T_c$ with 
the experimentally determined transition temperature 
$T^\ast$. Since now both $N$ and $T_c$ are fixed directly 
by experimental data, 
there is only one remaining adjustable parameter, namely $\nu$. 
According to our discussion above, the fit is expected to yield 
values $\nu \approx 1$.


In Figs.~\ref{figR2} and \ref{figR4} the experimental curves are fitted 
with Eq.~(\ref{resis}) using $\nu$ as a fit parameter. 
We obtain as optimal fit values $\nu=0.75$ for sample $R2$ 
and $\nu=0.16$ for sample $R4$, respectively. 
The first is in very good agreement with the expected theoretical 
value of $\nu$. For sample $R4$, the optimal $\nu$ is smaller than expected, 
which  may indicate that dissipative processes are more important in that sample. 
It is also possible that the  screening of 
Coulomb interactions is less effective in this narrow rope (where nearly half 
of the tubes are on the surface) than in the thicker rope $R2$.
Nevertheless, for both samples, the low-temperature 
resistance agrees quite well, with only one free fit parameter that 
is found to be of order unity as expected. 
The theoretical curves clearly do not provide 
a good description in the vicinity of $T_c$. This is however 
the expected consequence of the perturbative nature of our calculation, 
which breaks down close to $T_c$ due to QPS proliferation. 
Thus the saturation observed experimentally around $T\approx T^\ast$ 
is not captured. 
Equation~(\ref{resis}) also predicts a vanishing linear resistance as 
$T\rightarrow 0$. A finite $T=0$ resistance is usually expected when, 
instead of (or in addition to) bound pairs of QPSs, one also considers single QPS events. 
However, the latter are expected to be important only for short 1D systems 
and have not been taken into account in our theory.  An order-of-magnitude
estimate indicates anyway that their contribution to the resistance is 
exponentially small due to the large $T=0$ value of $\mu$, and then practically 
unmeasurable.

We believe that the rather good agreement between the theoretical 
resistance result (\ref{resis}) and experimental data 
at low temperatures as shown in Figs.~\ref{figR2} and \ref{figR4}, 
given the complexity of this system, is rather satisfactory. 
This comparison provides strong evidence that 
quantum phase slips have been observed in superconducting nanotube ropes. 
\acknowledgments
This work has been supported by the EU network DIENOW. 
A.K. thanks the Russian foundation for basic research and 
solid state nanostructures for financial support, and thanks 
CNRS for a visitor's position. We thank M. Devoret, N. Dupuis, 
T. Giamarchi, J. Gonz\'alez, D. Maslov, and C. Pasquier 
for stimulating discussions.

\end{document}